\documentclass[a4paper,11pt]{article}
\usepackage{amssymb,rotating,graphics,epsfig,float,graphics}
\usepackage{amsmath,verbatim,a4wide}
\usepackage[english]{babel}
\usepackage{amsmath,amsfonts,amssymb,pifont}
\usepackage{textcomp}
\usepackage{latexsym}
\usepackage{theorem,srcltx}
\usepackage{graphicx,float}
\usepackage{mathpazo} 
\usepackage{longtable}
\usepackage{txfonts}

\usepackage{theorem}
\theoremstyle{break} 
{\theorembodyfont{\rmfamily} }
{\theorembodyfont{\rmfamily} }
{\theorembodyfont{\rmfamily} }
{\theorembodyfont{\rmfamily} }

\begin{document}

\author{Roelof Bruggeman and Ferdinand Verhulst\\
University of Utrecht, Department of mathematics \\
PO Box 80.010, 3508 TA Utrecht, The Netherlands } 

\title{Optical-acoustical interaction in FPU-chains with alternating large masses (plus appendix)}

\maketitle
\begin{abstract} 
One of the problems of periodic FPU-chains with alternating masses is whether significant interactions exist between the
so-called (high frequency) optical and (low frequency) acoustical groups. We show that for $\alpha$-chains with 
$4n$ and $8n$ particles ($n= 1, 2, \ldots$) we have significant interactions caused by external 
forcing. For $\beta$-chains with 
$4$ and $8$ particles the interactions are characterised by parametric forcing, the interactions are negligible. 
\end{abstract} 

MSC classes:	37J20, 37J40, 34C20, 58K70, 37G05, 70H33, 70K30, 70K45\\


\section{Introduction} 
This note studies the alternating periodic FPU-chain (formulated in  \cite{GGMV}) for the special case of a large difference 
between the masses; it is a continuation of \cite{BValt}. The spectrum 
in this case can be divided into two groups, the optical group with relatively large frequencies, and the acoustical group 
with small frequencies.  
We will show that energy exchange can occur between the optical group and the acoustical group. As the system is 
Hamiltonian such exchange will always be reversible. 
\medskip

The spatially periodic FPU-chain with $N$ particles where the first oscillator is connected with the last one can be described
by the Hamiltonian 
 \begin{equation} \label{Hfpu}
 H(p, q) = \sum_{j=1}^N \left( \frac{1}{2m_j}p_j^2 + V(q_{j+1} - q_j)\right), 
 \end{equation} 

We choose the number $N=2n$ of particles even and take the odd masses $m_{2j+1}$ equal to $1$, the much larger even masses 
$m_{2j}=\frac{1}{a}$, where in this note $a>0$ is small. The chain is called an alternating FPU-chain. 

We consider the Hamiltonian near stable equilibrium $p=q=0$, and use a potential $V$ of the form 
\[ V(z) = \frac{1}{2} z^2+\frac{\alpha}{3} z^3+\frac{\beta}{4}z^4\,,\]
and speak of an $\alpha$-chain if $\alpha \neq 0,\beta=0$ and of a $\beta$-chain if $\alpha =0, \beta \neq 0$. 

We mention some facts concerning such systems, and refer to  our paper \cite{BValt} for more details. The eigenvalues of the linear 
system are given  in [3, Proposition 3.2]. With $2n$ particles in the chain the optical group is characterised by $n$ eigenvalues of size 
$2+O(a)$ and   
we have $n$ eigenvalues of size $O(a)$, the acoustical group, which includes the eigenvalue $0$ corresponding to the
momentum integral.

If the number $2n$ of particles is a $4$-fold, the system has the eigenvalues $2(1+a)$, $2$ (in the optical  group), and $2a$ (in the 
acoustical group).  For each of these eigenvalues we found explicit periodic solutions, for the $\alpha$-chain as well as the $\beta$-chain. 
(See section 2 of \cite{BValt}.)

Theorem 3.1 in \cite{BValt} is an embedding result. It states that each alternating periodic FPU-chain with $2n \geq 4$ particles 
occurs isomorphically as an invariant submanifold in all subsystems with $2kn$ particles ($k\in \mathbb{Z}_{\geq 2}$) with the same 
parameters $a$, $\alpha$, and $\beta$. So the study of small alternating FPU-chains is relevant for larger alternating systems.\\ 

 A question posed in \cite{GGMV} is whether there is energy exchange between the two groups or, formulated differently, 
 can high frequency modes transfer energy to low frequency modes and vice versa? 
 Our answer will be affirmative for $\alpha$-chains. \\
 The parameters $\alpha, \beta$ scale the nonlinearities. To avoid the solutions of the acoustical group leaving the 
 domain of stable equilibrium around the origin, we can choose the initial values small or adjust  $\alpha, \beta$. The 
 techniques are equivalent. The interaction dynamics will be characterised by using actions $E(t)$ (the traditional form 
 of the quadratic part of the energies), defined at the end of 
 next section, or by computing the Euclidean distance to the initial values as a function of time. 

 \section{Periodic FPU $\alpha$-chain with 4  particles} \label{sectionalfa4} 
 
\begin{figure}[ht]
\begin{center}
\includegraphics[width=5cm]{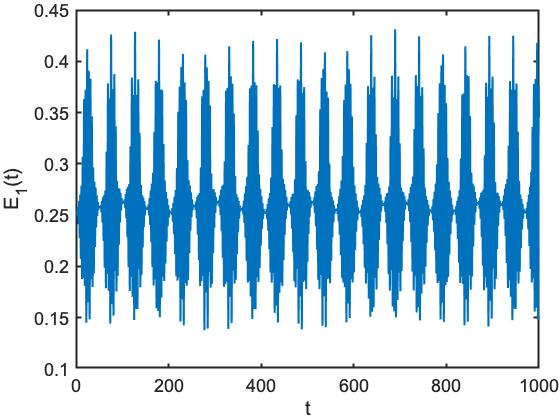}\,\includegraphics[width=5cm]{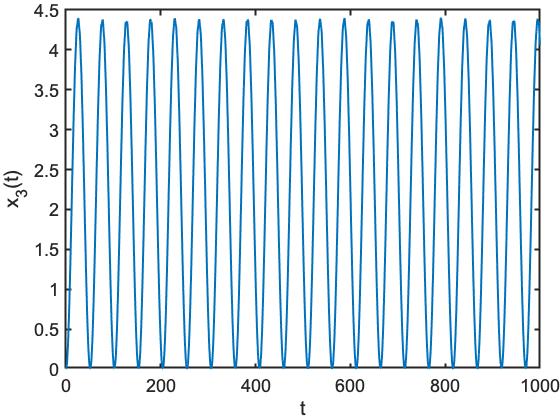}\,\includegraphics[width=5cm]{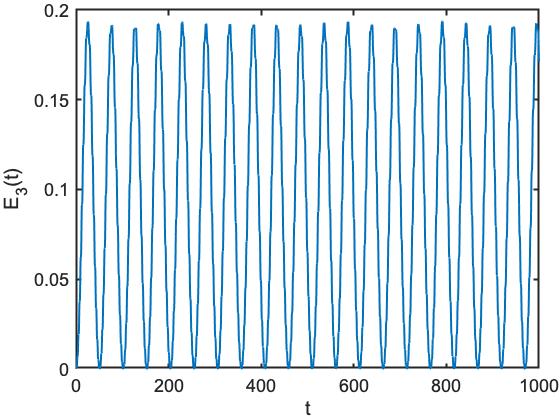}
\end{center}
 \caption{We illustrate the interaction of the optical and acoustical groups  of system \eqref{alfa4}  for $a= 0.01$. 
 Initial positions $x_1(0) =x_2(0)=0.5, x_3(0)=0$ and initial velocities zero.
Left $E_1(t)$ (optical group), middle $x_3(t)$ and right $E_3(t)$ (acoustical group) (the scales for $E_1$ and $E_3$ 
are different).}
\label{alfa4inter}
\end{figure} 

Systems with 4 particles are embedded in systems with $4n$ particles. 
In the case of 4 particles 
 we find the eigenvalues $2(1+a), 2, 2a, 0$. Reduction to 3 degrees-of-freedom (dof) produces for the $\alpha$-chain 
 the system: 
 \begin{eqnarray} \label{alfa4}
\begin{cases}
\ddot{x}_1 + 2(1+a) x_1 & =  2 \alpha \sqrt{a(1+a)} x_2 x_3,\\ 
\ddot{x}_2 + 2 x_2 & =  2 \alpha \sqrt{a(1+a)} x_1 x_3,\\
\ddot{x}_3 + 2a x_3 & =  2 \alpha \sqrt{a(1+a)} x_1 x_2. 
\end{cases} 
\end{eqnarray} 
We choose here $\alpha =1$ and the initial values suitable. 
For system \eqref{alfa4} the 3 normal modes exist and are harmonic functions with frequencies 
\[ \sqrt{2(1+a)}, 2, 2a. \] 
This problem for $a$ small was studied in \cite{FV20} with the conclusion that there exist strong interaction between the 
optical group (modes 1 and 2) and the acoustical group (mode 3) when starting in a neighbourhood of periodic 
solutions of the detuned resonance formed by the modes 1 and 2. (When starting at initial conditions that are not close 
to a periodic solution in $1:1$ resonance for modes $x_1, x_2$, the interaction is neglible.) 
The interaction is demonstrated in \cite{FV20} by constructing normal forms for system \eqref{alfa4}.
The asymptotic approximation to $O(\sqrt{a})$ obtained in this way leads to a forced, linear 
equation for $x_3(t)$: 
\begin{equation} \label{slowx3} 
\ddot{x}_3 + 2a x_3  =  2 \sqrt{a} r_0^2 \cos^2 (\sqrt{2}t + \psi_0), 
\end{equation} 
$\psi_0$ is the constant phase of the optical periodic solution, $r_0$ the amplitude. 
The general solution with constants $c_1, c_2$ is: 
\begin{equation} \label{acous}
x_3(t)= \frac{r_0^2}{2 \sqrt{a}}  - \frac{r_0^2}{8r_0^2 - 2 \sqrt{a}}\cos (2 \sqrt{2}t +2 \psi_0)+ c_1 \cos(\sqrt{2a}t) 
+ c_2 \sin(\sqrt{2a}t). 
\end{equation}

We illustrate the interaction in fig.~\ref{alfa4inter} with the actions 
$E_1(t)= 0.5(\dot{x}_1^2+ 2(1+a) x_1^2), E_3(t)= 0.5(\dot{x}_3^2+ 2a x_5^2)$. The interaction is spectacular. 

\section{Periodic FPU $\beta$-chain with 4 particles} \label{sec4}
 In this case the reduction to 3 dof, choosing $\beta =1$, becomes for 4 particles (see  \cite{BV18}): 
 \begin{eqnarray} \label{beta4} 
\begin{cases}
\ddot{x}_1 + 2(1+a) x_1 & =  -x_1(x_1^2+3x_2^2)  -ax_1(2x_1^2+3x_2^2+3x_3^2) -a^2x_1(x_1^2+3x_3^2),\\ 
\ddot{x}_2 + 2 x_2 & =  -x_2(x_2^2+3x_1^2) -3ax_2(x_1^2+x_3^2),\\
\ddot{x}_3 + 2a x_3 & =  -3ax_3(x_1^2+x_2^2)- a^2 x_3 (x_3^2+3x_1^2). 
\end{cases} 
\end{eqnarray} 
The detuned $1:1$ resonance of the modes 1 and 2 can be studied by averaging-normalisation in the usual way 
($x_1 \mapsto \varepsilon x_1, \dot{x}_1 \mapsto \varepsilon \dot{x}_1$ etc, see \cite{BV18}); this leads to in-phase 
and out-phase periodic 
solutions that can be approximated shortly by putting $x_1(t)= \pm x_2(t)$. To $O(a)$ we find for these periodic solutions:
\begin{equation} \label{perbeta4}
\ddot{x}_1 + 2x_1= -4x_1^3, \, x_1(t) = \pm x_2(t). 
\end{equation} 
The equation for $x_3$ becomes in this approximation:
\[ \ddot{x}_3 + 2a(1+ 3x_1^2(t))x_3=0, \] 
with $x_1(t)$ a periodic solution. For the Floquet exponents we have $\lambda_1 + \lambda_2=0$. \\ 
In the 4 particles $\alpha$-chain the $x_3$ mode with small frequency is excited by a periodic, semi-definite forcing, whereas for the 
4 particles $\beta$-chain we have parametric excitation with widely different freqencies, close to $\sqrt{2a}$ and $\sqrt{2}$. Even if these 
frequencies are resonant, the Floquet instability tongues will be extremely narrow for $a \rightarrow 0$. 
Interactions between optical and acoustical modes are then negligible for the 4 particles $\beta$-chain. Numerical 
experiments confirm this. 

 \section{Periodic FPU $\alpha$-chain with 8  particles} \label{sec4}
 Systems with 8 particles are imbedded in systems with $8n$ particles.  
 We use system (23) and Table 1 of \cite{BValt} to determine the 7 dof system describing the alternating FU-chain with 
 8 particles. The eigenvalues of the linearised system are: 
 \[ 2(a+1), \sqrt{a^2+1} +a+1\,{\rm (twice)}, 2, 2a, -\sqrt{a^2+1}+a+1\,{\rm (twice)}, 0. \] 
 To show the relative size of the terms in the equations of motion we include for the optical group (equations 1-4) the 
 nonlinear terms to $O(\sqrt{a})$. We find for this group: 
  \begin{eqnarray} \label{alfa8opt} 
\begin{cases}
\ddot{x}_1 + 2 x_1 & =   \alpha  \sqrt{a}(2x_4x_5  +\sqrt{2}x_3x_6   +\sqrt{2}x_2x_7) + O(a^{\frac{3}{2}}),\\ 
\ddot{x}_2 + 2 x_2 & =  \alpha  \sqrt{a} ( \sqrt{2}x_7x_4+x_2x_5 + \sqrt{2}x_1x_7) + O(a^{\frac{3}{2}}),\\
\ddot{x}_3 + 2 x_3 & =    \alpha  \sqrt{a}(- \sqrt{2}x_6x_4 - x_3x_5 +\sqrt{2} x_1x_6) + O(a^{\frac{3}{2}}),\\ 
 \ddot{x}_4 + 2 x_4 & =   \alpha  \sqrt{a}(- \sqrt{2}x_6x_3 + 2x_1x_5 +\sqrt{2} x_2x_7) + O(a^{\frac{3}{2}}). 
 \end{cases} 
 \end{eqnarray}
 In the same way the acoustical group with  nonlinear terms to $O(\sqrt{a})$ becomes: 
  \begin{eqnarray} \label{alfa8ac} 
\begin{cases}
\ddot{x}_5 +2a x_5 & = \alpha  \sqrt{a}(x_2^2 -x_3^2  + 2x_1x_4) + O(a^{\frac{3}{2}}),\\

\ddot{x}_6 + a x_6 & =  \alpha  \sqrt{2a} (x_1x_3 - x_4x_3 ) + O(a^{\frac{3}{2}}),\\

\ddot{x}_7 + a x_7 & =   \alpha  \sqrt{2a}( x_1x_2 + x_4x_2)  + O(a^{\frac{3}{2}}). 
 
 \end{cases} 
 \end{eqnarray}
For the $\alpha$-chain with 8 particles  3 invariant manifolds were found in  \cite{BValt}. First, we consider 
possible interactions between optical and acoustical modes in these manifolds. We 
choose $\alpha =1$ and small initial values.\\

{\bf Invariant manifold $M_{145}$}\\ 
This manifold with modes 1, 4, 5 corresponds with the case of 4 particles described by system \eqref{alfa4} 
in section \ref{sectionalfa4}. 
The embedding of the manifold $M_{145}$ was formulated and proved in \cite[Theorem 3.1]{BValt}. 
It shows interactions between optical and acoustical modes as the periodic solutions of the resonant modes 1 and 4 
are forcing mode 5.\\

{\bf Invariant manifold $M_{256}$}\\ 

\begin{figure}[ht]
\begin{center}
\includegraphics[width=6cm]{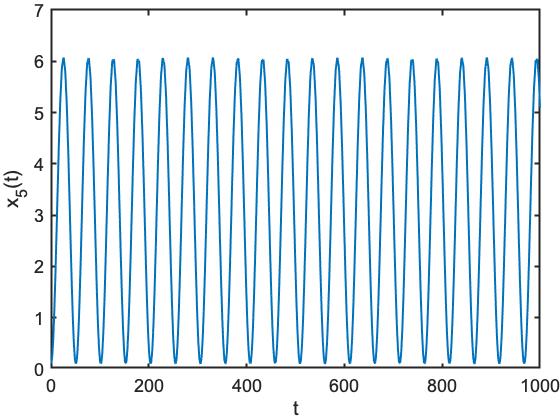}\,\includegraphics[width=6cm]{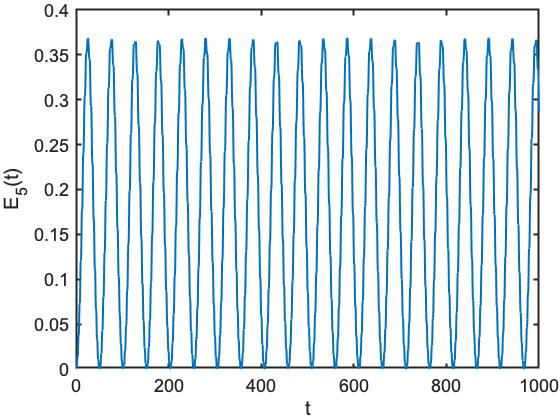}
\end{center}
 \caption{Left we illustrate the behaviour of the solutions  of system \eqref{alfa256} describing the dynamics in 
 invariant manifold $M_{256}$. Initial positions $x_2(0) =1, x_5(0)=0.1, x_6(0)=0.1$ and initial velocities zero; $a= 0.01$.
Left $x_5(t)$ and right $E_5(t)= 0.5(\dot{x}_5^2+ 2a x_5^2)$. }
\label{IM256exc}
\end{figure} 
For the equations of motion we need the following coefficients:
\begin{equation*}
d_{225}= - \sqrt{a}\frac{\sqrt{a^2+1} +a+1}{\sqrt{a^2+1}},\,d_{256}=\frac{\sqrt{2}a^2}{\sqrt{a^2+1}},\,
d_{566}= \sqrt{a}   \frac{- \sqrt{a^2+1} +a+1}{2 \sqrt{a^2+1}}.
\end{equation*}

 The dynamics in the manifold $M_{256}$ consisting of modes 2, 5, 6 is described by the system: 
 \begin{eqnarray} \label{alfa256} 
\begin{cases}
\ddot{x}_2 + (\sqrt{a^2+1} +a+1) x_2 & =    - 2 d_{225}x_2x_5 - d_{256}x_5x_6,\\ 

\ddot{x}_5 +2a x_5 & =  -d_{225}x_2^2 -  d_{256}x_2x_6 -  d_{566}x_6^2,\\

\ddot{x}_6 + (-\sqrt{a^2+1}+a+1) x_6 & =  -d_{256}x_2x_5 - 2d_{566}x_5x_6. 
 \end{cases} 
 \end{eqnarray} 
 The only (exact) normal mode periodic solution is found for $x_5$ which is harmonic. In this manifold the optical group counts 1 mode ($x_2$), 
 the acoustical group 2 modes. The external forcing of mode $x_5$ by mode $x_2$  is clear from system \eqref{alfa256}. 
 See fig.~\ref{IM256exc} where the excitation of mode 5 is shown, $x_6$ is not excited.\\

{\bf Invariant manifold $M_{357}$}\\
 The dynamics in the manifold $M_{357}$ consisting of modes 3, 5, 7 shows similar behaviour as in manifold $_{256}$.
 We have  from \cite{BValt}:
 \begin{equation*} 
 d_{355} = -d_{225},\, d_{357}= d_{256},\, d_{775}=  - d_{566}. 
 \end{equation*}
  It is described 
 by the system of the form: 
 \begin{eqnarray} \label{alfa357} 
\begin{cases}
\ddot{x}_3 + (\sqrt{a^2+1} +a+1) x_3 & =    - 2d_{335}x_3x_5 - d_{357}x_5x_7),\\ 

\ddot{x}_5 +2a x_5 & =  -d_{335}x_3^2 - d_{775} x_7^2 - d_{357}x_3x_7,\\

\ddot{x}_7 + (-\sqrt{a^2+1}+a+1) x_7 & =  - d_{357}x_3x_5 - 2d_{775}x_5x_7. 
 \end{cases} 
 \end{eqnarray} 

  For $a \rightarrow 0$  we have  
 $d_{335} = O(\sqrt{a}), d_{357}= O(a^2), d_{775} = O(a^{\frac{5}{2}})$. There is one optical mode, 2 acoustical ones. 
 Again mode $x_5$ is excited, in this case by mode $x_3$; $x_7$ shows very small variations. The behaviour is similar to the 
 dynamics in manifold $M_{256}$.\\
 
\begin{figure}[ht]
\begin{center}
\includegraphics[width=5cm]{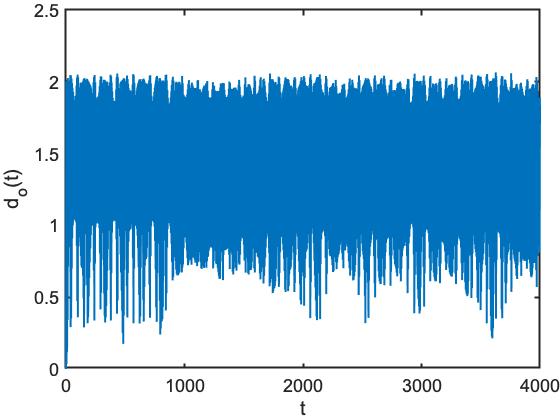}\,\includegraphics[width=5cm]{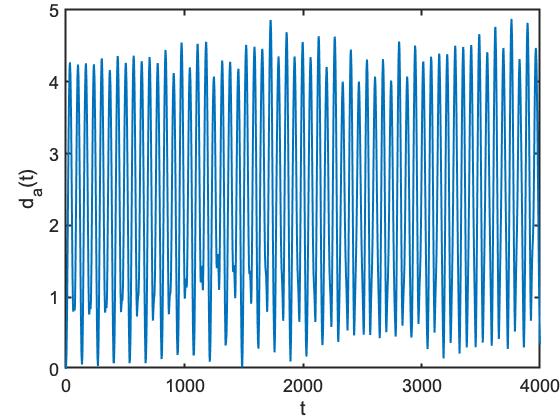}\,\includegraphics[width=5cm]{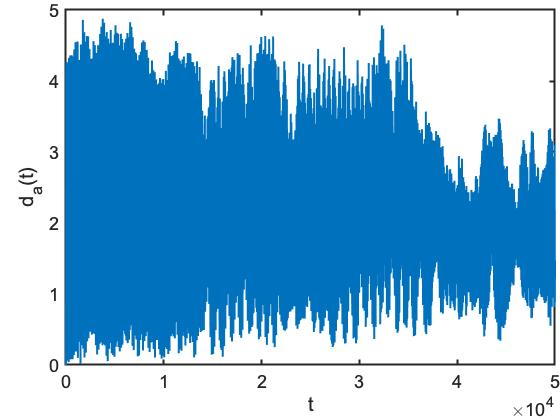}
\end{center}
 \caption{We illustrate for $a= 0.01$ the interaction of the optical and acoustical groups  of system (23) in \cite{BValt}  
 (derived for $\alpha$-chains with 8 particles). 
 Initial positions $x_j(0)=0.5, j= 1, \ldots, 4$,, $x_j(0)=0, j= 5, 6, 7$ and initial velocities zero. Left $d_o(t)$ for 
 4000 timesteps. 
The acoustical group is mainly excited by $x_5(t)$ (middle figure shows $d_a(t)$ on 4000 timesteps), as an 
illustration $d_a(t)$ is also shown for $50\,000$ timesteps.} 
\label{alfa7general}
\end{figure}

{ \bf General initial conditions}\\ 
 As we can see by inspection of the Hamiltonian with coefficients given in table 1 of \cite{BValt} there are more terms 
 leading to interaction of the optical and acoustical group, for instance the terms $x_3x_4x_6$, $x_2x_4x_7$. We 
 illustrate the interaction by choosing nonzero initial values in the optical group and zero initial values in the 
 acoustical group, see fig.~\ref{alfa7general}. We use the distance to the initial values of the optical and acoustical groups:
  \begin{equation} \label{distdo}
 d_o(t)= \sqrt{\sum_{j=1}^4 ((x_j(t)-x_j(0))^2 +  (\dot{x}_j(t)-\dot{x}_j(0))^2)}, 
\end{equation} 
 and 
   \begin{equation} \label{distda}
 d_a(t)= \sqrt{\sum_{j=5}^7 ((x_j(t)-x_j(0))^2 +  (\dot{x}_j(t)-\dot{x}_j(0))^2)}. 
 \end{equation}
 On a time interval of 1000 steps the interaction is dominated by $x_5(t)$; for a more complete picture we also give 
 $d_a(t)$ for $50\,000$ timesteps in fig.~\ref{alfa7general}. 
 
 We conclude that in periodic $\alpha$-chains with 8n particles 
 we have {\em significant interaction between the optical and the acoustical groups}. In each case mode $x_5$ plays an important part.

  \section{Periodic FPU $\beta$-chain with 8 particles} \label{sec5}
  Using \cite{BValt}   and the appendix in section \ref{sec7} we can write down the cubic part of the 7 dof Hamiltonian; it contains 49 terms 
  dependent on $a$. As discussed before the invariant manifold $M_{145}$ corresponds with the description for 4 particles 
  in section \ref{sec4}. At this stage we keep the choice of the parameter $\beta$ open. \\ 
  
  {\bf Invariant manifold $M_{256}$}\\
  The equations of motion for modes 2, 5 and 6 to $O(a^{\frac{3}{2}})$ are: 
   \begin{eqnarray} \label{beta256} 
\begin{cases}
\ddot{x}_2 + (\sqrt{a^2+1} +a+1) x_2 & =   - \beta(  (1+a)x_3^3+\frac{3}{2}ax_3x_5^2)  
+O(a^{\frac{3}{2}}),\\ 

\ddot{x}_5 +2a x_5 & = - \beta \frac{3}{2}ax_2^2x_5+O(a^{\frac{3}{2}}),\\

\ddot{x}_6 + (-\sqrt{a^2+1}+a+1) x_6 & =  O(a^{\frac{3}{2}}). 
 \end{cases} 
 \end{eqnarray} 
 In the limit for $a \rightarrow 0$ we find periodic solutions from the $x_2$ limit equation: 
 \begin{equation} \label{8partx2beta} 
 \ddot{x}_2 + 2 x_2= - \beta x_2^3. 
 \end{equation} 
 If $\beta >0$ the solutions of the limit equation are all periodic, if $\beta <0$ the solutions are periodic in an $O(1)$ 
 neighbourhood of the origin. The $O(a)$ perturbation term modulates the period on a long timescale $O(1/ a^{\frac{3}{2}})$. 
 For $x_5$ we have the parametrically excited equation: 
 \begin{equation} \label{8partx5beta} 
 \ddot{x}_5 + 2 a(1+ \beta \tfrac{3}{4}x_2^2)= O(a^{\frac{3}{2}}). 
 \end{equation} 
 If $\beta >0$ we expect (as in the case of $M_{145}$) only small interaction between optical and acoustical modes, 
 if $\beta <0$ the same conclusion holds for small values of $\beta$ or in a neighbourhood of the origin. \\

{\bf Invariant manifold $M_{357}$}\\
For modes 3, 5 and 7 in invariant manifold  $M_{357}$ we find to $O(a^{\frac{3}{2}})$:  
 \begin{eqnarray} \label{beta357} 
\begin{cases}
\ddot{x}_3 + (\sqrt{a^2+1} +a+1) x_3 & =   - \beta(  (1+a)x_2^3+\frac{3}{2}ax_2x_5^2)  
+O(a^{\frac{3}{2}}),\\ 

\ddot{x}_5 +2a x_5 & = - \beta \frac{3}{2}ax_2^2x_5+O(a^{\frac{3}{2}}),\\

\ddot{x}_7 + (-\sqrt{a^2+1}+a+1) x_7 & =  O(a^{\frac{3}{2}}). 
 \end{cases} 
 \end{eqnarray} 
 The equations are analogous to the system for $M_{236}$, the discussion runs along the same lines.\\
 {\em We conclude that for 8 particles $\beta$-chains no significant interaction between optical and acoustical 
 modes takes place in the 3 invariant manifolds.}\\
 
 {\bf General initial conditions}\\ 
 This is a more complicated case as among the 49 coefficients of the Hamiltonian we have terms like 
 $x_1x_2x_3x_5$,  $x_1x_2x_3x_5$, $x_1x_3x_4x_7$; in the equations of motion these terms will produce a certain 
 forcing of modes $x_5, x_6, x_7$. However, in all the cases of such forcing terms they have coefficients that are 
 $O(a^{3/2})$,  they will have less influence. See also the discussion in section \ref{sec6}. 
  
 \section{Conclusions} \label{sec6} 

\newcommand\CC{{\mathbb C}}

For alternating periodic FPU-chains with a large
mass ratio, we showed that for $\alpha$-chains with 4 (or $4n$) and 8 (or $8n$)
particles that a strong interaction may occur between optical and
acoustic frequencies. A special part is played by the eigenmode with
eigenvalue $2a$, which occurs in systems with $4n$ and $8n$ modes. 
In systems with 4 or 8 particles we find the eigencoordinates with eigenvalues: 
\[ \begin{array}{c|cc|c}
2n &2(1+a) & 2 & 2a\\
\hline
4&x_1&x_2&x_3\\
8&x_1&x_4&x_5
\end{array}\]
In the case of 8 particles there is an additional optical eigenvalue
$1+a+\sqrt{1+a^2}$ with multiplicity~$2$, corresponding to the
eigencoordinates $x_2$ and $x_3$.

For the $\alpha$-chains the external forcing is visible in the system by
the presence in the quadratic contribution in system \eqref{alfa8ac}
 to $\ddot x_5$ of three terms $x_1 x_4$, $x_2^2 $ and $x_3^2$,
(the term with $x_1x_4$ corresponds to $x_1x_2$ in system 
\eqref{alfa4} in the line with $\ddot x_3$. Suitable
starting values of the optical eigencoordinates have a strong influence
on the eigenmode with eigenvalue $2a$.\\

For the $\beta$-chains we look at the cubic contribution to $\ddot x_3$
in \eqref{beta4}, which is of the form $x_3$ times a
negative definite expression of size $\mathrm{O}(a)$. For $8$ particles
the quadratic contribution to $\ddot x_5$ has the form
\begin{equation}
-\frac{3a}2 x_5 \bigl( x_1^2+x_2^2+x_3^2+x_4^2\bigr)
-\frac{3a}{\sqrt2}\bigl( -x_1x_3x_6+x_3x_4x_6+x_1x_2x_7+x_2x_4x_7\bigr)
+ \mathrm{O}(a^{3/2})\,.
\end{equation}
This does not give rise to external forcing.\medskip

Theorem~3.1 in \cite{BValt} implies that an alternating periodic
FPU-chain with $2n$ particles is present as an invariant submanifold
in all alternating periodic FPU-chains with $2nk$ particles for all
$k \geq 2$, with the same parameters $a$, $\alpha$ and $\beta$. We have
seen this explicitly for the systems with 4 particles arising as the invariant
submanifold $M_{145}$  in the system with $8$ particles.

Two questions arise:
\begin{enumerate}
\item Are there more optical eigenmodes in an $\alpha$-chain
with $4n$ particles that influences the eigenmode for the eigenvalue
$2a$ by external forcing?
\item Can there be optical eigenmodes in a $\beta$-chain with
$4n$ particles that influence the eigenmode for the eigenvalue $2a$ by
external forcing?
\end{enumerate}

Question 1 for $\alpha$-chains has the answer \emph{yes}. 
The embedding result
\cite[Theorem 3.1]{BValt} is proved by a decomposition
$\CC^{2} = \bigoplus_\zeta X(\zeta)$ with two-dimensional subspace
$X(\zeta)$ parametrised by $n$-th roots of unity~$\zeta$. The
eigenspaces with eigenvalues $2a$ and $0$ are contained in $X(1)$, the
eigenspaces with eigenvalues $2$ and $2a$ in $X(-1)$. For all other
$n$-th roots of unity the space $X(\zeta)$ contains an eigenvector in
the optical group and an eigenvector in the acoustic group that are
both non-real. To get real eigenvectors we have to consider the sum
$X(\zeta) \oplus X(\bar\zeta)$.

Let $\eta_{-1}$ be the coordinate of the eigenvector with eigenvalue
$2a$, contained in $X(-1)$. (So $\eta_{-1}$ is proportional to $x_3$ in
the system with 4 particles, and proportional to $x_5$ in the system
with 8 particles.)
The quadratic contribution to $\ddot \eta_{-1}$ is obtained from
products of eigencoordinates corresponding to subspaces $X(\zeta_1) $
and $X(\zeta_2)$ such that $\zeta_1\zeta_2=-1$; see \cite[(16)]{BValt}.
Working this out we get products of an optical eigencoordinate related
to $X(\zeta_1) \oplus X(\overline{\zeta_1})$ and an optical
eigencoordinate related to $X(\zeta_2)\oplus X(\overline{\zeta_2})$,
and more quadratic monomials in which at most one factor is an optical
eigencoordinate. So, for $\alpha$-chains with $4n$ particles we get optical contributions like
$x_1x_4$ in \eqref{alfa8ac}. Only if $\zeta_1=\zeta_2 $
is equal to $i$ or $-i$, squares of optical eigencoordinates are
possible. This correspond to the squares $x_2^2$ and $x_3^2$ in \eqref{alfa8ac}.

\begin{figure}[ht]
\begin{center}
\includegraphics[width=7cm]{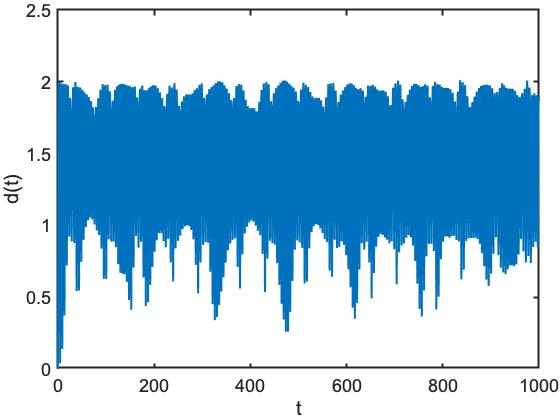}\,\includegraphics[width=7cm]{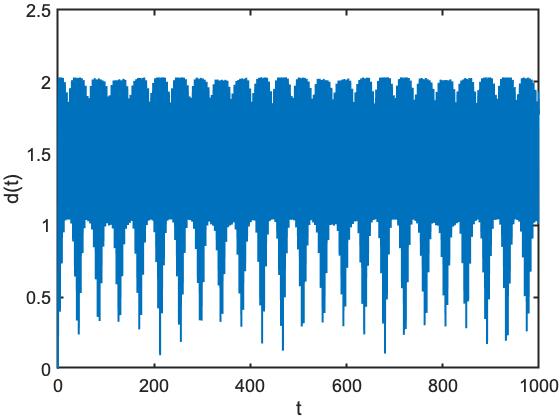}
\end{center}
 \caption{Left the Euclidean distance for the $\alpha$-chain, right the $\beta$-chain, both with 8 particles 
 (based on Hamiltonian \eqref{Hfpu} with $N=8, a= 0.01$). In both cases the initial conditions correspond 
 with optical position $0.5$, acoustical positions $0$ and all initial velocities zero; we took respectively for 
 $\alpha$ and $\beta$ the value $0.5$. 
 Both cases show instability of the optical group but this is weaker and shows more regular recurrence 
 in the case of the $\beta$-chain. }
\label{dist8part}
\end{figure} 
\medskip

Question 2 is more difficult. For $\beta$-chains with 4 or 8 particles we found no significant interaction,
see fig.\ref{dist8part} for an illustration of the time series of the Euclidean distance to the initial values in 
an $\alpha$-chain and a $\beta$-chain with 8 particles. 
Note that an optical group does not represent an invariant manifold but we expect that strong interaction 
with the acoustical group will increase and complicate the recurrence in the system expressed by the Euclidean 
distance.\\
More in general:  
if $x_1, x_2, x_3$ are eigenmodes of the optical group and 
$x_a$ an eigenmode of the acoustical group, the Hamiltonian will in general contain terms 
of the form $x_1x_2x_3x_a$. If we have for $x_1, x_2, x_3$ solutions that are periodic or  
nearly periodic, this may result in forcing of the acoustical mode. As we discussed before, in the 
case of 8 particles such terms are $O(a^{3/2})$ i.e dominated by other nonlinear terms. 
The mathematics of this interaction  can be exemplified by the cartoon equation for acoustical modes 
(see also the coefficients in the appendix, section \ref{sec7}):
\begin{equation} \label{cartoon}
 \ddot{x} + a x = a \beta (\cos^2(2t) x + \surd a \cos^3(2t)), 
 \end{equation}  
where $a= 1/m$ is a small parameter. Here we have replaced the optical modes by $\cos(2t)$ which correponds 
with period $2+ o(a)$ solutions that satisfy the optical system, see system \eqref{alfa8opt}.
If $\beta \ll 1$ or altenatively we rescale the coordinates 
close to the origin of phase-space, we have again a parametrically excited equation with extremely 
narrow instability tongues. In this case we conclude that for $\beta$ small enough or at low energy values 
we expect no significant interaction between optical and acoustical groups.

\section{Appendix}  \label{sec7}
For the case of 8 particles the coefficients used in the systems of section \ref{sec4} can be found in \cite{BValt} table 1. 
The $\beta$-chain of the alternating FPU-chain discussed in section \ref{sec5} has the form

{\begin{center}
\begin{longtable}{rl}
$\ddot x_1 $&= $-2 (a+1) x_1$\\&
$\quad\hbox{}-\beta\Bigl(4 x_1^3 e_{1,1,1,1}
+2 x_1 x_2^2 e_{1,1,2,2}
+2 x_1 x_3^2 e_{1,1,3,3}
+x_2^2 x_4 e_{1,2,2,4}
$\\
&$\quad\hbox{}+x_3^2 x_4 e_{1,3,3,4}
+2 x_1 x_4^2 e_{1,1,4,4}
+x_2 x_3 x_5 e_{1,2,3,5}
+2 x_1 x_5^2 e_{1,1,5,5}
$\\
&$\quad\hbox{}+x_2 x_4 x_6 e_{1,2,4,6}
+x_3 x_5 x_6 e_{1,3,5,6}
+2 x_1 x_6^2 e_{1,1,6,6}
+x_4 x_6^2 e_{1,4,6,6}
$\\
&$\quad\hbox{}+x_3 x_4 x_7 e_{1,3,4,7}
+x_2 x_5 x_7 e_{1,2,5,7}
+x_5 x_6 x_7 e_{1,5,6,7}
+2 x_1 x_7^2 e_{1,1,7,7}
$\\
&$\quad\hbox{}+x_4 x_7^2 e_{1,4,7,7}
\Bigr)$\\ 
$\ddot x_2 $&= $-\bigl(\sqrt{a^2+1}+a+1\bigr) x_2$\\&
$\quad\hbox{}-\beta\Bigl(2 x_1^2 x_2 e_{1,1,2,2}
+4 x_2^3 e_{2,2,2,2}
+2 x_2 x_3^2 e_{2,2,3,3}
+2 x_1 x_2 x_4 e_{1,2,2,4}
$\\
&$\quad\hbox{}+2 x_2 x_4^2 e_{2,2,4,4}
+x_1 x_3 x_5 e_{1,2,3,5}
+2 x_2 x_5^2 e_{2,2,5,5}
+3 x_2^2 x_6 e_{2,2,2,6}
$\\
&$\quad\hbox{}+x_3^2 x_6 e_{2,3,3,6}
+x_1 x_4 x_6 e_{1,2,4,6}
+2 x_2 x_6^2 e_{2,2,6,6}
+x_6^3 e_{2,6,6,6}
$\\
&$\quad\hbox{}+2 x_2 x_3 x_7 e_{2,2,3,7}
+x_1 x_5 x_7 e_{1,2,5,7}
+x_4 x_5 x_7 e_{2,4,5,7}
+x_3 x_6 x_7 e_{2,3,6,7}
$\\
&$\quad\hbox{}+2 x_2 x_7^2 e_{2,2,7,7}
+x_6 x_7^2 e_{2,6,7,7}
\Bigr)$\\ 
$\ddot x_3 $&= $-\bigl(\sqrt{a^2+1}+a+1\bigr) x_3$\\&
$\quad\hbox{}-\beta\Bigl(2 x_1^2 x_3 e_{1,1,3,3}
+2 x_2^2 x_3 e_{2,2,3,3}
+4 x_3^3 e_{3,3,3,3}
+2 x_1 x_3 x_4 e_{1,3,3,4}
$\\
&$\quad\hbox{}+2 x_3 x_4^2 e_{3,3,4,4}
+x_1 x_2 x_5 e_{1,2,3,5}
+2 x_3 x_5^2 e_{3,3,5,5}
+2 x_2 x_3 x_6 e_{2,3,3,6}
$\\
&$\quad\hbox{}+x_1 x_5 x_6 e_{1,3,5,6}
+x_4 x_5 x_6 e_{3,4,5,6}
+2 x_3 x_6^2 e_{3,3,6,6}
+x_2^2 x_7 e_{2,2,3,7}
$\\
&$\quad\hbox{}+3 x_3^2 x_7 e_{3,3,3,7}
+x_1 x_4 x_7 e_{1,3,4,7}
+x_2 x_6 x_7 e_{2,3,6,7}
+x_6^2 x_7 e_{3,6,6,7}
$\\
&$\quad\hbox{}+2 x_3 x_7^2 e_{3,3,7,7}
+x_7^3 e_{3,7,7,7}
\Bigr)$\\ 
$\ddot x_4 $&= $-2 x_4$\\&
$\quad\hbox{}-\beta\Bigl(x_1 x_2^2 e_{1,2,2,4}
+x_1 x_3^2 e_{1,3,3,4}
+2 x_1^2 x_4 e_{1,1,4,4}
+2 x_2^2 x_4 e_{2,2,4,4}
$\\
&$\quad\hbox{}+2 x_3^2 x_4 e_{3,3,4,4}
+4 x_4^3 e_{4,4,4,4}
+2 x_4 x_5^2 e_{4,4,5,5}
+x_1 x_2 x_6 e_{1,2,4,6}
$\\
&$\quad\hbox{}+x_3 x_5 x_6 e_{3,4,5,6}
+x_1 x_6^2 e_{1,4,6,6}
+2 x_4 x_6^2 e_{4,4,6,6}
+x_1 x_3 x_7 e_{1,3,4,7}
$\\
&$\quad\hbox{}+x_2 x_5 x_7 e_{2,4,5,7}
+x_1 x_7^2 e_{1,4,7,7}
+2 x_4 x_7^2 e_{4,4,7,7}
\Bigr)$\\ 
$\ddot x_5 $&= $-2 a x_5$\\&
$\quad\hbox{}-\beta\Bigl(x_1 x_2 x_3 e_{1,2,3,5}
+2 x_1^2 x_5 e_{1,1,5,5}
+2 x_2^2 x_5 e_{2,2,5,5}
+2 x_3^2 x_5 e_{3,3,5,5}
$\\
&$\quad\hbox{}+2 x_4^2 x_5 e_{4,4,5,5}
+4 x_5^3 e_{5,5,5,5}
+x_1 x_3 x_6 e_{1,3,5,6}
+x_3 x_4 x_6 e_{3,4,5,6}
$\\
&$\quad\hbox{}+2 x_5 x_6^2 e_{5,5,6,6}
+x_1 x_2 x_7 e_{1,2,5,7}
+x_2 x_4 x_7 e_{2,4,5,7}
+x_1 x_6 x_7 e_{1,5,6,7}
$\\
&$\quad\hbox{}+2 x_5 x_7^2 e_{5,5,7,7}
\Bigr)$\\ 
$\ddot x_6 $&= $-\bigl(-\sqrt{a^2+1}+a+1\bigr) x_6$\\&
$\quad\hbox{}-\beta\Bigl(x_2^3 e_{2,2,2,6}
+x_2 x_3^2 e_{2,3,3,6}
+x_1 x_2 x_4 e_{1,2,4,6}
+x_1 x_3 x_5 e_{1,3,5,6}
$\\
&$\quad\hbox{}+x_3 x_4 x_5 e_{3,4,5,6}
+2 x_1^2 x_6 e_{1,1,6,6}
+2 x_2^2 x_6 e_{2,2,6,6}
+2 x_3^2 x_6 e_{3,3,6,6}
$\\
&$\quad\hbox{}+2 x_1 x_4 x_6 e_{1,4,6,6}
+2 x_4^2 x_6 e_{4,4,6,6}
+2 x_5^2 x_6 e_{5,5,6,6}
+3 x_2 x_6^2 e_{2,6,6,6}
$\\
&$\quad\hbox{}+4 x_6^3 e_{6,6,6,6}
+x_2 x_3 x_7 e_{2,3,6,7}
+x_1 x_5 x_7 e_{1,5,6,7}
+2 x_3 x_6 x_7 e_{3,6,6,7}
$\\
&$\quad\hbox{}+x_2 x_7^2 e_{2,6,7,7}
+2 x_6 x_7^2 e_{6,6,7,7}
\Bigr)$\\ 
$\ddot x_7 $&= $-\bigl(-\sqrt{a^2+1}+a+1\bigr) x_7$\\&
$\quad\hbox{}-\beta\Bigl(x_2^2 x_3 e_{2,2,3,7}
+x_3^3 e_{3,3,3,7}
+x_1 x_3 x_4 e_{1,3,4,7}
+x_1 x_2 x_5 e_{1,2,5,7}
$\\
&$\quad\hbox{}+x_2 x_4 x_5 e_{2,4,5,7}
+x_2 x_3 x_6 e_{2,3,6,7}
+x_1 x_5 x_6 e_{1,5,6,7}
+x_3 x_6^2 e_{3,6,6,7}
$\\
&$\quad\hbox{}+2 x_1^2 x_7 e_{1,1,7,7}
+2 x_2^2 x_7 e_{2,2,7,7}
+2 x_3^2 x_7 e_{3,3,7,7}
+2 x_1 x_4 x_7 e_{1,4,7,7}
$\\
&$\quad\hbox{}+2 x_4^2 x_7 e_{4,4,7,7}
+2 x_5^2 x_7 e_{5,5,7,7}
+2 x_2 x_6 x_7 e_{2,6,7,7}
+2 x_6^2 x_7 e_{6,6,7,7}
$\\
&$\quad\hbox{}+3 x_3 x_7^2 e_{3,7,7,7}
+4 x_7^3 e_{7,7,7,7}
\Bigr)$
\end{longtable}
\end{center}}

The following table gives the coefficients.
{\small\begin{center}
\begin{longtable}{rlrl}
$e_{1111}$&
$\;=\;\frac{1}{8} (a+1)^2$&
$\;=\;$&$\frac{1}{8}+\frac{a}{4}+\frac{a^2}{8}+O\left(a^3\right)$\\
$e_{1122}$&
$\;=\;\frac{3}{8} (a+1) \left(\sqrt{a^2+1}+a+1\right)$&
$\;=\;$&$\frac{3}{4}+\frac{9 a}{8}+\frac{9 a^2}{16}+O\left(a^3\right)$\\
$e_{1133}$&
$\;=\;\frac{3}{8} (a+1) \left(\sqrt{a^2+1}+a+1\right)$&
$\;=\;$&$\frac{3}{4}+\frac{9 a}{8}+\frac{9 a^2}{16}+O\left(a^3\right)$\\
$e_{1144}$&
$\;=\;\frac{3 (a+1)}{4}$&
$\;=\;$&$\frac{3}{4}+\frac{3 a}{4}+O\left(a^3\right)$\\
$e_{1155}$&
$\;=\;\frac{3}{4} a (a+1)$&
$\;=\;$&$\frac{3 a}{4}+\frac{3 a^2}{4}+O\left(a^3\right)$\\
$e_{1166}$&
$\;=\;-\frac{3}{8} (a+1) \left(\sqrt{a^2+1}-a-1\right)$&
$\;=\;$&$\frac{3 a}{8}+\frac{3 a^2}{16}+O\left(a^3\right)$\\
$e_{1177}$&
$\;=\;-\frac{3}{8} (a+1) \left(\sqrt{a^2+1}-a-1\right)$&
$\;=\;$&$\frac{3 a}{8}+\frac{3 a^2}{16}+O\left(a^3\right)$\\
$e_{1224}$&
$\;=\;\frac{3 \sqrt{a+1} \left(\sqrt{a^2+1}+a+1\right)}{4 \
\sqrt{a^2+1}}$&
$\;=\;$&$\frac{3}{2}+\frac{3 a}{2}-\frac{3 a^2}{16}+O\left(a^3\right)$\\
$e_{1235}$&
$\;=\;-\frac{3 a^{3/2} \sqrt{a+1} \left(\sqrt{a^2+1}+a+1\right)}{2 \
\sqrt{a^2+1}}$&
$\;=\;$&$-3 a^{3/2}-3 a^{5/2}+O\left(a^3\right)$\\
$e_{1246}$&
$\;=\;-\frac{3 a^{3/2} \sqrt{a+1}}{\sqrt{2} \sqrt{a^2+1}}$&
$\;=\;$&$-\frac{3 a^{3/2}}{\sqrt{2}}-\frac{3 a^{5/2}}{2 \
\sqrt{2}}+O\left(a^3\right)$\\
$e_{1257}$&
$\;=\;\frac{3 a \sqrt{a+1}}{\sqrt{2} \sqrt{a^2+1}}$&
$\;=\;$&$\frac{3 a}{\sqrt{2}}+\frac{3 a^2}{2 \sqrt{2}}+O\left(a^3\right)$\\
$e_{1334}$&
$\;=\;-\frac{3 \sqrt{a+1} \left(\sqrt{a^2+1}+a+1\right)}{4 \
\sqrt{a^2+1}}$&
$\;=\;$&$-\frac{3}{2}-\frac{3 a}{2}+\frac{3 a^2}{16}+O\left(a^3\right)$\\
$e_{1347}$&
$\;=\;-\frac{3 a^{3/2} \sqrt{a+1}}{\sqrt{2} \sqrt{a^2+1}}$&
$\;=\;$&$-\frac{3 a^{3/2}}{\sqrt{2}}-\frac{3 a^{5/2}}{2 \
\sqrt{2}}+O\left(a^3\right)$\\
$e_{1356}$&
$\;=\;-\frac{3 a \sqrt{a+1}}{\sqrt{2} \sqrt{a^2+1}}$&
$\;=\;$&$-\frac{3 a}{\sqrt{2}}-\frac{3 a^2}{2 \sqrt{2}}+O\left(a^3\right)$\\
$e_{1466}$&
$\;=\;-\frac{3 \sqrt{a+1} \left(-\sqrt{a^2+1}+a+1\right)}{4 \
\sqrt{a^2+1}}$&
$\;=\;$&$-\frac{3 a}{4}+O\left(a^3\right)$\\
$e_{1477}$&
$\;=\;\frac{3 \sqrt{a+1} \left(-\sqrt{a^2+1}+a+1\right)}{4 \
\sqrt{a^2+1}}$&
$\;=\;$&$\frac{3 a}{4}+O\left(a^3\right)$\\
$e_{1567}$&
$\;=\;-\frac{3 a^{3/2} \sqrt{a+1} \left(-\sqrt{a^2+1}+a+1\right)}{2 \
\sqrt{a^2+1}}$&
$\;=\;$&$-\frac{3 a^{5/2}}{2}+O\left(a^3\right)$\\
$e_{2222}$&
$\;=\;\frac{\left(a^2+2\right) \left(a^2+2 \sqrt{a^2+1}+2\right)}{8 \
\left(a^2+1\right) \left(\sqrt{a^2+1}-a+1\right)^2}$&
$\;=\;$&$\frac{1}{4}+\frac{a}{4}+\frac{a^2}{16}+O\left(a^3\right)$\\
$e_{2226}$&
$\;=\;\frac{a^{3/2} \left(a \left(\sqrt{a^2+1}-a+1\right)-2 \
\left(\sqrt{a^2+1}+1\right)\right)}{2 \sqrt{2} \left(a^2+1\right) \
\left(\sqrt{a^2+1}-a+1\right)^2}$&
$\;=\;$&$-\frac{a^{3/2}}{2 \sqrt{2}}-\frac{a^{5/2}}{4 \
\sqrt{2}}+O\left(a^3\right)$\\
$e_{2233}$&
$\;=\;\frac{3 a^2 \left(\sqrt{a^2+1}+1\right)^2}{4 \left(a^2+1\right) \
\left(\sqrt{a^2+1}-a+1\right)^2}$&
$\;=\;$&$\frac{3 a^2}{4}+O\left(a^3\right)$\\
$e_{2237}$&
$\;=\;-\frac{3 a^{3/2} \left(\sqrt{a^2+1}+1\right)}{2 \sqrt{2} \
\left(a^2+1\right) \left(\sqrt{a^2+1}-a+1\right)}$&
$\;=\;$&$-\frac{3 a^{3/2}}{2 \sqrt{2}}-\frac{3 a^{5/2}}{4 \
\sqrt{2}}+O\left(a^3\right)$\\
$e_{2244}$&
$\;=\;\frac{3}{8} \left(\sqrt{a^2+1}+a+1\right)$&
$\;=\;$&$\frac{3}{4}+\frac{3 a}{8}+\frac{3 a^2}{16}+O\left(a^3\right)$\\
$e_{2255}$&
$\;=\;\frac{3}{8} a \left(\sqrt{a^2+1}+a+1\right)$&
$\;=\;$&$\frac{3 a}{4}+\frac{3 a^2}{8}+O\left(a^3\right)$\\
$e_{2266}$&
$\;=\;\frac{3 a^3}{8 a^2+8}$&
$\;=\;$&$\frac{3a^2}8+O\left(a^5\right)$\\
$e_{2277}$&
$\;=\;\frac{3 a \left(a^2+2\right)}{8 \left(a^2+1\right)}$&
$\;=\;$&$\frac{3 a}{4}+O\left(a^3\right)$\\
$e_{2336}$&
$\;=\;\frac{3 a^{3/2} \left(\sqrt{a^2+1}+1\right)}{2 \sqrt{2} \
\left(a^2+1\right) \left(\sqrt{a^2+1}-a+1\right)}$&
$\;=\;$&$\frac{3 a^{3/2}}{2 \sqrt{2}}+\frac{3 a^{5/2}}{4 \
\sqrt{2}}+O\left(a^3\right)$\\
$e_{2367}$&
$\;=\;\frac{3 a^3}{2 a^2+2}$&
$\;=\;$&$ \frac{3a^3}2+O\left(a^5\right)$\\
$e_{2457}$&
$\;=\;\frac{3 a}{\sqrt{2}}$&
$\;=\;$&$\frac{3 a}{\sqrt{2}}+O\left(a^3\right)$\\
$e_{2666}$&
$\;=\;\frac{a^{5/2} \left(a \left(-2 \sqrt{a^2+1}+2 a-1\right)+
\sqrt{a^2+1}+1\right)}{2 \sqrt{2} \left(a^2+1\right) \
\left(\sqrt{a^2+1}-a+1\right)^2}$&
$\;=\;$&$\frac{a^{5/2}}{4 \sqrt{2}}+O\left(a^3\right)$\\
$e_{2677}$&
$\;=\;\frac{3 a^{5/2} \left(a-\sqrt{a^2+1}\right)}{2 \sqrt{2} \
\left(a^2+1\right) \left(\sqrt{a^2+1}-a+1\right)}$&
$\;=\;$&$-\frac{3 a^{5/2}}{4 \sqrt{2}}+O\left(a^3\right)$\\
$e_{3333}$&
$\;=\;\frac{\left(a^2+2\right) \left(a^2+2 \sqrt{a^2+1}+2\right)}{8 \
\left(a^2+1\right) \left(\sqrt{a^2+1}-a+1\right)^2}$&
$\;=\;$&$\frac{1}{4}+\frac{a}{4}+\frac{a^2}{16}+O\left(a^3\right)$\\
$e_{3337}$&
$\;=\;\frac{a^{3/2} \left(a^2-\left(\sqrt{a^2+1}+1\right) a+2 \
\left(\sqrt{a^2+1}+1\right)\right)}{2 \sqrt{2} \left(a^2+1\right) \
\left(\sqrt{a^2+1}-a+1\right)^2}$&
$\;=\;$&$\frac{a^{3/2}}{2 \sqrt{2}}+\frac{a^{5/2}}{4 \
\sqrt{2}}+O\left(a^3\right)$\\
$e_{3344}$&
$\;=\;\frac{3}{8} \left(\sqrt{a^2+1}+a+1\right)$&
$\;=\;$&$\frac{3}{4}+\frac{3 a}{8}+\frac{3 a^2}{16}+O\left(a^3\right)$\\
$e_{3355}$&
$\;=\;\frac{3}{8} a \left(\sqrt{a^2+1}+a+1\right)$&
$\;=\;$&$\frac{3 a}{4}+\frac{3 a^2}{8}+O\left(a^3\right)$\\
$e_{3366}$&
$\;=\;\frac{3 a \left(a^2+2\right)}{8 \left(a^2+1\right)}$&
$\;=\;$&$\frac{3 a}{4}+O\left(a^3\right)$\\
$e_{3377}$&
$\;=\;\frac{3 a^3}{8 a^2+8}$&
$\;=\;$&$\frac{3a^3}8+O\left(a^5\right)$\\
$e_{3456}$&
$\;=\;\frac{3 a}{\sqrt{2}}$&
$\;=\;$&$\frac{3 a}{\sqrt{2}}+O\left(a^3\right)$\\
$e_{3667}$&
$\;=\;\frac{3 a^{5/2} \left(\sqrt{a^2+1}-a\right)}{2 \sqrt{2} \
\left(a^2+1\right) \left(\sqrt{a^2+1}-a+1\right)}$&
$\;=\;$&$\frac{3 a^{5/2}}{4 \sqrt{2}}+O\left(a^3\right)$\\
$e_{3777}$&
$\;=\;-\frac{a^{5/2} \left(a \left(-2 \sqrt{a^2+1}+2 a-1\right)
+\sqrt{a^2+1}+1\right)}{2 \sqrt{2} \left(a^2+1\right) \
\left(\sqrt{a^2+1}-a+1\right)^2}$&
$\;=\;$&$-\frac{a^{5/2}}{4 \sqrt{2}}+O\left(a^3\right)$\\
$e_{4444}$&
$\;=\;\frac{1}{8}$&
$\;=\;$&$\frac{1}{8}+O\left(a^3\right)$\\
$e_{4455}$&
$\;=\;\frac{3 a}{4}$&
$\;=\;$&$\frac{3 a}{4}+O\left(a^3\right)$\\
$e_{4466}$&
$\;=\;-\frac{3}{8} \left(\sqrt{a^2+1}-a-1\right)$&
$\;=\;$&$\frac{3 a}{8}-\frac{3 a^2}{16}+O\left(a^3\right)$\\
$e_{4477}$&
$\;=\;-\frac{3}{8} \left(\sqrt{a^2+1}-a-1\right)$&
$\;=\;$&$\frac{3 a}{8}-\frac{3 a^2}{16}+O\left(a^3\right)$\\
$e_{5555}$&
$\;=\;\frac{a^2}{8}$&
$\;=\;$&$\frac{a^2}{8}+O\left(a^3\right)$\\
$e_{5566}$&
$\;=\;\frac{3}{8} a \left(-\sqrt{a^2+1}+a+1\right)$&
$\;=\;$&$\frac{3 a^2}{8}+O\left(a^3\right)$\\
$e_{5577}$&
$\;=\;\frac{3}{8} a \left(-\sqrt{a^2+1}+a+1\right)$&
$\;=\;$&$\frac{3 a^2}{8}+O\left(a^3\right)$\\
$e_{6666}$&
$\;=\;\frac{a^2 \left(a^2+2\right) \left(2 a \left(a-\sqrt{a^2+1}\right)
+1\right)}{8 \left(a^2+1\right) \left(\sqrt{a^2+1}-a+1\right)^2}$&
$\;=\;$&$\frac{a^2}{16}+O\left(a^3\right)$\\
$e_{6677}$&
$\;=\;\frac{3 a^2 \left(a^2-\sqrt{a^2+1} a-\sqrt{a^2+1}+a+1\right)}{8 \
\left(a^2+1\right)}$&
$\;=\;$&$\frac{3a^4}{16}+
O\left(a^5\right)$\\
$e_{7777}$&
$\;=\;\frac{a^2 \left(a^2+2\right) \left(2 a \left(a-\sqrt{a^2+1}\right)
+1\right)}{8 \left(a^2+1\right) \left(\sqrt{a^2+1}-a+1\right)^2}$&
$\;=\;$&$\frac{a^2}{16}+O\left(a^3\right)$\
\end{longtable}
\end{center}}


\begin{thebibliography}{99} 

\bibitem{BV18} R.W. Bruggeman and F. Verhulst, {\em Dynamics of a chain with four particles and nearest- neighbor 
interaction}, in Recent Trends in Applied Nonlinear Mechanics and Physics (M. Belhaq, ed.), CSNDD 2016, pp. 103-120, 
doi 10.1007/978-3-319-63937-6-6, Springer (2018).


\bibitem{BValt} Roelof Bruggeman and Ferdinand Verhulst, {\em Near-integrability and recurrence in FPU 
chains with alternating masses}, J. Nonlinear Science 29, pp. 183-206, DOI 10.1007/s00332-018-9482-x (2019).


\bibitem{GGMV} L. Galgani, A. Giorgilli, A. Martinoli and S. Vanzini, On the problem of energy partition for large 
systems of the Fermi-Pasta-Ulam type: analytical and numerical estimates, Physica D 59, pp. 334-348 (1992).  


\bibitem{FV20} Ferdinand Verhulst, {\em  Variations on the Fermi-Pasta-Ulam chain, a survey}, arXiv: 
2003.09156 (2020), submitted to Chaotic Modeling and Simulations (CMSIM).


\end{thebibliography}
\end{document}